\documentclass[floatfix, preprint, onecolumn, prd,
showpacs, showkeys, preprintnumbers, nofootinbib, superscriptaddress]{revtex4-1}
\usepackage{natbib}
\usepackage{epstopdf}
\usepackage{times}
\usepackage{amssymb,amsbsy,amsmath,amsfonts}
\usepackage{graphicx}
\usepackage{float}
\usepackage{color}
\usepackage{morefloats}
\usepackage{rotating}
\usepackage{srcltx}
\usepackage{slashed}
\usepackage{subfigure}
\usepackage{multirow}
\usepackage{verbatim}
\usepackage{hyperref}
\usepackage{tabularx}
\usepackage{epstopdf}

\graphicspath{{./}{./img/}{./fig/}{./image/}{./figure/}{./picture/}}

\begin{document}

\title{Test of the $h_1(1830)$ made of $K^{*}\bar{K}^*$ with the $\eta_c\rightarrow \phi K^{*}\bar{K}^{*}$ decay}

\author{Xiu-Lei Ren}
\affiliation{School of Physics and
Nuclear Energy Engineering \& International Research Center for Nuclei and Particles in the Cosmos, Beihang University, Beijing 100191, China}

\author{Li-Sheng Geng}
\email[E-mail: ]{lisheng.geng@buaa.edu.cn}
\affiliation{School of Physics and
Nuclear Energy Engineering \& International Research Center for Nuclei and Particles in the Cosmos, Beihang University, Beijing 100191, China}

\author{E. Oset}
\email[E-mail: ]{oset@ific.uv.es}
\affiliation{School of Physics and
Nuclear Energy Engineering \& International Research Center for Nuclei and Particles in the Cosmos, Beihang University, Beijing 100191, China}
\affiliation{Departamento de F\'{\i}sica Te\'{o}rica and IFIC,
Centro Mixto Universidad de Valencia-CSIC, Institutos de Investigaci\'{o}n de Paterna,
Apartado 22085, 46071 Valencia, Spain}

\author{Jie Meng}
\affiliation{School of Physics and
Nuclear Energy Engineering \& International Research Center for Nuclei and Particles in the Cosmos, Beihang University, Beijing 100191, China}
\affiliation{State Key Laboratory of Nuclear Physics and Technology, School of Physics, Peking University, Beijing 100871, China}
\affiliation{Department of Physics, University of Stellenbosch, Stellenbosch 7602, South Africa}

\begin{abstract}
  We present a new reaction, complementary to the $J/\psi\rightarrow \eta K^{*0}\bar{K}^{*0}$ from which an $h_1$ resonance with mass around $1830$ MeV was reported from a BESIII experiment. The new reaction is $\eta_c\rightarrow \phi K^*\bar{K}^*$, or $\eta_c(2S)\rightarrow \phi K^*\bar{K}^*$. Using the information from the analysis of $J/\psi\rightarrow \eta K^{*0}\bar{K}^{*0}$, we find that the $K^*\bar{K}^*$ invariant mass distribution for those two $\eta_c$ decays exhibits a clear peak around $1830$ MeV perfectly distinguishable from what one obtains with pure phase space. We suggest the implementation of these reactions to assert the existence of this elusive resonance which, by its nature as a vector-vector molecule with $0^-(1^{+-})$ quantum numbers, only couples to the $K^*\bar{K}^*$ channel.
\end{abstract}


\date{\today}

\maketitle
\section{Introduction}
The advent of experimental facilities, BES, Belle, BaBar, CLEO~\cite{Ali:2011vy,Gersabeck:2012rp,Olsen:2012zz,Li:2012pd} has brought an impressive advance in the field of Hadron Physics and has discovered a great deal of new states in the light quark sector, and with open and hidden charm or beauty~\cite{SANTEL:2013jua,Olsen:2014mea}. Mesonic states have been the most favored, usually observed through peaks in the invariant mass of pairs of particles, trios or even bigger numbers of more elementary mesonic states. The interpretation of the results requires to deal with the meson-meson interaction and there, the chiral unitary theory (UCPT) \cite{Kaiser:1995eg,npa,review}, or extensions of it using the local hidden gauge approach~\cite{hidden1,hidden2,hidden4}, have proved very efficient. One of the issues in hadronic physics is to learn about the structure of the states found, and in this sense the conventional wisdom of mesons made of $q \bar q$ and baryons of $qqq$ has been challenged by both experiment and theory \cite{Klempt:2007cp,Klempt:2009pi,Brambilla:2010cs}. Indeed, some meson states clearly demand tetraquark structures or a molecular description, while some baryonic states clearly call for more complex structures, like the $\Lambda(1405)$, which for long was claimed to correspond to a quasibound $\bar K N$ state \cite{Dalitz:1960du}. The pseudoscalar meson-meson interaction has been extensively studied in UCPT \cite{npa,Kaiser:1998fi,Markushin:2000fa,Dobado:1996ps,Pelaez:2006nj}.  In many cases, the interaction, smoothly energy dependent, is sufficient to produce bound states or resonances when coupled channels are used. These states are referred as dynamically generated in the Literature. They arise as poles of the scattering matrix from the solution of the Bethe-Salpeter equation in coupled channels, using the potential provided by the tree level amplitudes of chiral perturbation theory or the local hidden gauge approach. Sometimes they are also referred as composite states, following the nomenclature of Weinberg in his early work that showed that the deuteron was an ordinary bound state of a neutron and a proton and not a genuine state of other nature with small coupling to $p$ and $n$~\cite{weincompo}.

The vector meson-vector meson interaction has been only addressed more recently \cite{raquel,gengvec,gengnieves,albaoller}.  The use of the local hidden gauge approach made possible the systematic study of this interaction, which started with the study of the $\rho \rho$ interaction~\cite{raquel} and was extended to the SU(3) space in Ref.~\cite{gengvec}. In Ref.~\cite{gengvec}, in addition to the $f_2(1270)$ and $f_0(1370)$ found in Ref.~\cite{raquel}, nine more resonances were found, most of which could be associated to known states, while a few remained as predictions. Among the predictions there is an $h_1 ~0^-(1^{+-})$ around 1800 MeV that couples to $K^* \bar K^*$ as  the only channel. The fact that ordinary decay channels of this state are not allowed, because of its quantum numbers and structure, has probably been a reason why this state is not catalogued in the PDG \cite{pdg}. Yet, a lucky coincidence, the measurement of the $J/\psi \to \eta K^{*0}\bar{K}^{*0}$ decay by the BES Collaboration~\cite{BESdata} searching for the $Y(2175)$, now catalogued as $\phi(2170)$ in the PDG (which incidentally was not found in~Ref.~\cite{BESdata}, see Ref.~\cite{albaxie} for explanations), showed evidence for this $h_1$ state. Indeed in that decay a pronounced peak was observed in the invariant mass of $K^{*0}\bar{K}^{*0}$ around $1830$ MeV that an analysis carried out in Ref.~\cite{albaxie} attributed to the formation of the $h_1$ resonance made from the $K^* \bar K^*$ interaction. The presence of a resonance produced a distribution drastically different than that of pure phase space and in good agreement with experiment. In view of this finding, it is important that new reactions are measured which provide further evidence of this new state, such that it can become one of the accepted states according to the standards of the PDG. This is the purpose of the present paper.

In this work, based on the information obtained from Refs.~\cite{BESdata,albaxie}, we propose to study the decay of $\eta_c\rightarrow \phi K^*\bar{K}^*$ (${K^{*}}^+{K^{*}}^-$ or $K^{*0}\bar{K}^{*0}$), 
focusing on the invariant mass distributions of the $K^{*}\bar{K}^{*}$ pair. We predict an invariant mass distribution which has a shape drastically different from the one we obtain using simple phase space. The measurement of this distribution should bring additional information that would serve to confirm the new state proposed in Ref.~\cite{albaxie} and enrich the field of hadronic resonances with one which is clearly composite of two vector mesons.

\section{Theoretical Framework}

\begin{figure*}[t]
  \centering
  \includegraphics[width=16cm]{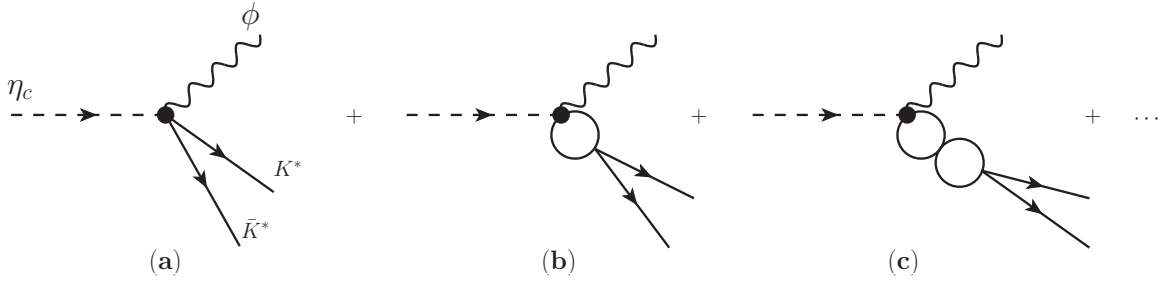}
  \caption{Diagrammatic representation of the $\eta_c\rightarrow\phi{K^{*}} \bar{K}^{*}$ decay. (a) tree level, (b) and (c) with final state interaction of the $K^{*}\bar{K}^*$. Production with $L=0$ is assumed in order 
  to attribute the quantum number of an $h_1$ to the $K^*\bar{K}^*$ state.}
  \label{Fig:FeynDiagrams}
\end{figure*}

The process $\eta_c\rightarrow \phi K^{*}\bar{K}^{*}$ can be depicted in Fig.~\ref{Fig:FeynDiagrams}. The diagram of Fig.~\ref{Fig:FeynDiagrams}-(a) shows the bare vertex for $\phi$, $K^*$, and $\bar{K}^*$ production. Diagrams Fig.~\ref{Fig:FeynDiagrams}-(b) and (c) account for rescattering of the $K^*$'s.

In diagrams Fig.~\ref{Fig:FeynDiagrams}-(b) and (c), the sum of terms after the first loop, including the tree level $K^*\bar{K}^*\rightarrow K^*\bar{K}^*$ interaction (bare vertex of the interaction of four $K^*$'s) and further iterations of it, can be summed with the Bethe-Salpeter equation to give the $K^*\bar{K}^*\rightarrow K^*\bar{K}^*$ $t$ matrix through
\begin{eqnarray}\label{Eq:tVG}
  t &=& V + VGV + VGVGV +\cdots \nonumber\\
    &=& V + V G t = V( 1 + G t) = \frac{V}{1 - V G} = \frac{1}{V^{-1} - G},
\end{eqnarray}
where $G$ is the loop function of the $K^*\bar{K}^*$ pair.

The simplest possibility, and chosen by nature unless forbidden by some symmetry, is that the process proceeds with $L=0$. This is also most natural since in this case there is not much phase space for the reaction ($M_{\eta_c}-m_{\phi}-2m_{K^*}=180$ MeV). Then the $K^*\bar{K}^*$ system is created, and propagates with the $I^G(J^{PC})=0^-(1^{+-})$ quantum numbers, those of an $h_1$ resonance. It is in this channel that the interaction provided by the local hidden gauge approach generates an $h_1$ structure around $1800$ MeV upon rescattering, according to Ref.~\cite{gengvec}. We shall not distinguish between charged or neutral $K^*$ since the $K^*\bar{K}^*$ state is created in $I=0$,
\begin{equation}
  |K^*\bar{K}^*,~I=0\rangle = -\frac{1}{\sqrt{2}}({K^*}^+{K^*}^- + {K^*}^0{\bar K}^{*0}),
\end{equation}
and propagates with this isospin. It suffices to project at the end over the ${K^*}^+{K^*}^-$ or ${K^*}^0{\bar K}^{*0}$ component, but since we are only interested in the shape of the distribution, this factor is unnecessary.

The $G$ function is known analytically in dimensional regularization~\cite{gengvec}, but since the $K^*$ has a width of about $50$ MeV, it is necessary to perform a convolution of the standard $G$ function to account for the mass distributions of the two $K^*$'s. This is done replacing $G$ by $\tilde{G}$ with
\begin{equation}
  \tilde{G}(s) = \frac{1}{\mathcal{N}^2}\int_{m_-^2}^{m_+^2} dm_1^2~dm_2^2~\left(-\frac{1}{\pi}\right)^2\omega(m_1^2)~\omega(m_2^2)~G(s,~m_1^2,~m_2^2),
\end{equation}
Here, $m_{1,2}$ represent the masses of two stable particles. The integration is in the range $m_{\pm}=m_K^*\pm 2\Gamma_{K^*}$ with the nominal mass $m_K^*$ and width $\Gamma_{K^*}$ of the $K^*$ meson, respectively.
The factor $\mathcal{N}$ is used for normalization and is given by
\begin{equation}
  \mathcal{N}^2 = \int_{m_{-}^2}^{m_+^2}dm_1^2dm_2^2\left(-\frac{1}{\pi}\right)^2\omega(m_1^2)\omega(m_2^2),
\end{equation}

\begin{equation}
  \omega(m^2) = \mathrm{Im}\frac{1}{m^2-m_{K^*}^2+i\Gamma(m^2)m},
\end{equation}

\begin{equation}
\Gamma(m^2)=\Gamma_{K^*}\frac{p^3(m^2)}{p^3(m_{K^*}^2)},\quad p(m^2)=\frac{\lambda^{1/2}(m^2,~m_\pi^2,~m_K^2)}{2m}\Theta(m-m_\pi-m_K).
\end{equation}

The standard $G$ loop function can be written as
\begin{eqnarray}
  G(s,~m_1^2,~m_2^2) &=& \frac{1}{16\pi^2}\left[a(\mu)+\log\frac{m_1m_2}{\mu^2} + \frac{\Delta}{2s}\log\frac{m_2^2}{m_1^2} \right.\nonumber\\
  && \left. + \frac{\nu}{2s}\left(\log\frac{s-\Delta+\nu}{-s+\Delta+\nu} + \log\frac{s+\Delta+\nu}{-s-\Delta+\nu}\right)\right],
\end{eqnarray}
with $\Delta=m_2^2-m_1^2$ and $\nu=\lambda^{1/2}(s,~m_1^2,~m_2^2)$, $\lambda(x,~y,~z)=(x-y-z)^2-4yz$, the K\"{a}hlen function.

If we call $t_P$ the bare production vertex of Fig.~\ref{Fig:FeynDiagrams}(a), the sum of all diagrams in the Fig.~\ref{Fig:FeynDiagrams} can be written as
\begin{equation}
  t_P = V_P + V_P \tilde{G} t = V_P(1+\tilde{G}t),
\end{equation}
and the factor $1+\tilde{G}t$ can be obtained from Eq.~(\ref{Eq:tVG}) as $t/V$.
Then we write
\begin{equation}
  t_P = V_P \frac{t}{V},
\end{equation}
and finally the $K^*\bar{K}^*$ invariant mass distribution of the $\eta_c\rightarrow \phi K^*\bar{K}^*$ can be written as
\begin{equation}\label{Eq:MinvSpec}
  \frac{d\Gamma}{d M_\mathrm{inv}} = \frac{C}{|V|^2}\frac{p_1\tilde{p}_2}{M_{\eta_c}^2}|t|^2,
\end{equation}
where $C$ is an arbitrary constant that absorbs $V_P$ and other constant factors, $p_1$ is the $\phi$ momentum in the $\eta_c$ rest frame,
\begin{equation}
  p_1 = \frac{\lambda^{1/2}(M_{\eta_c}^2,~m_\phi^2,~M_\mathrm{inv}^2)}{2M_{\eta_c}},
\end{equation}
and $\tilde{p}_2$ is the momentum of the $K^*$ in the $K^*\bar{K}^*$ center mass frame, $p_2$,
\begin{equation}
  p_2 = \frac{\lambda^{1/2}(M_\mathrm{inv}^2,~m_1^2,~m_2^2)}{2M_\mathrm{inv}}\Theta(M_\mathrm{inv}-m_1-m_2),
\end{equation}
convoluted by the mass distributions of the two $K^*$, as has been done for the $G$ function.

We take the input to construct $t$ from Ref.~\cite{albaxie}, which was fitted to the BES data~\cite{BESdata}. The potential $V$ is taken from Ref.~\cite{gengvec} and $\mu=1$ GeV. The subtraction constant $a(\mu)=-1.0$ is taken from~Ref.~\cite{albaxie} which reproduced well the data.

\section{Results and Discussion}

\begin{figure}[h!]
  \centering
  \includegraphics[width=10cm]{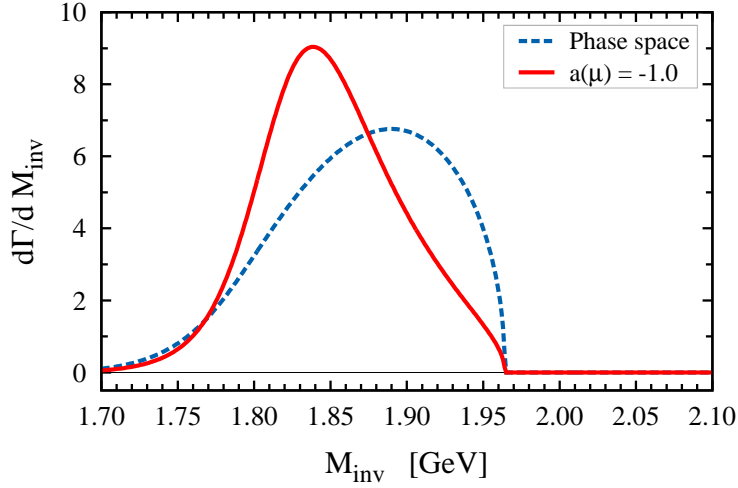}\\
  \caption{(Color online). The $K^{*}\bar{K}^{*}$ invariant mass spectrum of $\eta_c\rightarrow\phi K^{*}\bar{K}^{*}$ decay. The red solid line represents the results obtained
  with $a(\mu)=-1.0$, and the blue dashed line is the phase space.}
  \label{Fig:Fig2}
\end{figure}

In Fig.~\ref{Fig:Fig2}, we show the results for $d\Gamma/d M_\mathrm{inv}$ obtained with arbitrary renormalization in all the range of $M_\mathrm{inv}$.

We can see that a peak develops for the distribution around $1830$ MeV, where the mass of the resonance was deduced in~\cite{albaxie}. One might think that the limited phase space for this decay forces a shape with a peak. This is why in order to establish the presence of a resonance, one must compare this with what one expects from pure phase space. This is done by taking $t/V=1$ in Eq.~(\ref{Eq:MinvSpec}) and normalizing to the same area $\int dM_\mathrm{inv} d\Gamma/dM_\mathrm{inv}$ as in the resonant case. The result is depicted in Fig.~\ref{Fig:Fig2}, and we can see that the shapes are indeed very different and clearly distinguishable in an experiment without the need of excessive precision. Note that the phase space distribution accumulates most of the strength at higher invariant masses, while in the case of the resonance a clear and narrow peak occurs at lower energies.

\begin{figure}[t!]
  \centering
  \includegraphics[width=10cm]{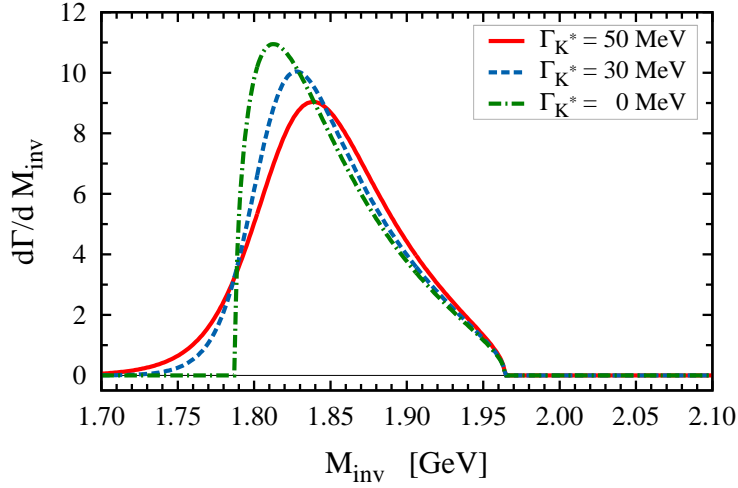}\\
  \caption{(Color online). The $K^{*}\bar{K}^{*}$ invariant mass spectrum of the $\eta_c\rightarrow \phi K^{*}\bar{K}^{*}$ decay with the same $a(\mu)=-1.0$ and different values of $\Gamma_{K^*}$. The solid red line corresponds to the physical case ($\Gamma_{K^*}=50$ MeV), the blue dashed line and green dot-dashed line stand for $\Gamma_{K^*}=30$ MeV and $\Gamma_{K^*}=0$ MeV, respectively.}
  \label{Fig:Fig3}
\end{figure}

It is worth seeing what happens if we make the width of the $K^*$ smaller. In Fig.~\ref{Fig:Fig3} we show the results assuming $\Gamma_{K^*}=50$ MeV, $30$ MeV or $0$ MeV. What we observe is that the peak of the distribution is shifed to lower invariant masses. Certainly, the results that we want are those with the actual width $\Gamma_{K^*}=50$ MeV, but we note that with the same input for $\mu$, $a(\mu)$, and $V$, the peak of the resonance is shifted to lower energies. In fact, if we take $\Gamma_{K^*}=0$ with that input, we find no pole but a pronounced cusp structure corresponding to a virtual state. This is in contrast with the case with $\Gamma_{K^*}=50$ MeV where the peak appears around $1830$ MeV. This is above the nominal threshold $2m_{K^*}=1780$ MeV and it is the mass distribution of the $K^*$ what makes the appearance above threshold possible. In fact, it was shown in~\cite{yamajuan} that a smooth potential independent of the energy cannot produce a resonance above threshold in a single channel.

\begin{figure}[b!]
  \centering
  \includegraphics[width=10cm]{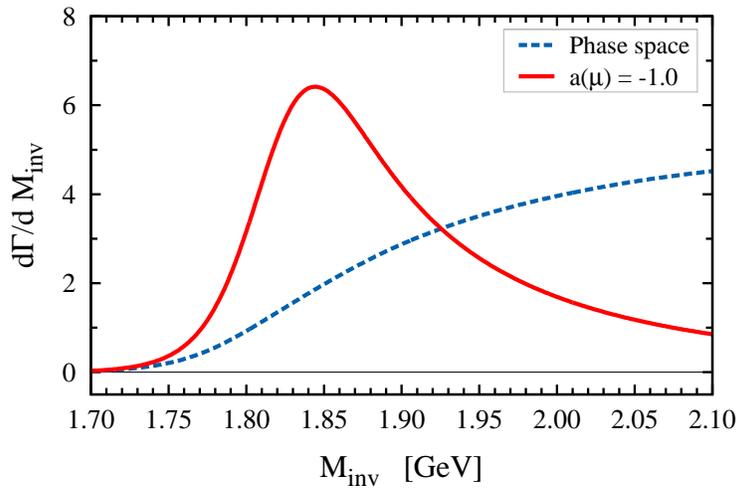}\\
  \caption{(Color online). The $K^{*+}K^{*-}$ invariant mass spectrum of $\eta_c(2S)\rightarrow\phi K^{*}\bar{K}^{*}$ decay. The red solid line represents the results obtained with
   $a(\mu)=-1.0$, and the blue dashed line is the  phase space.}
  \label{Fig:Fig4}
\end{figure}

The argument given here can be easily extended to the $\eta_{c}(2S)$ (or any other radially excited state of the $\eta_c$), since the quantum numbers are the same. The case of the $\eta_c(2S)$ provides a larger phase space for the decay and one might find different shapes. We have also performed the calculations for this case and the results are shown in Fig.~\ref{Fig:Fig4}.

We take a restricted region of $M_\mathrm{inv}$, from threshold to $2.1$ GeV where our results are reliable, and once again plot the results obtained with the $h_1$ resonance or just phase space, normalized to the same area. We see now results similar to those found in~\cite{albaxie}. The resonance case shows a clear peak in this region, while the phase space keeps rising in all the invariant mass range.

Clearly these two shapes would be perfectly distinguishable in an experiment.

\section{Conclusions}

We have studied the $\eta_c\rightarrow \phi {K^{*}}{\bar K^{*}}$ decay channel and have evaluated the shape of the invariant mass distribution taking into account the final state interaction of the ${K^{*}}{\bar K^{*}}$ system. For small phase space, or in any case relatively close to threshold, the s-wave production should be dominant and this fixes the quantum numbers of the  ${K^{*}}{\bar K^{*}}$ system to $0^- (1^{+-})$, those of an $h_1$ state.  In these quantum numbers the ${K^{*}}{\bar K^{*}}$ interaction is attractive and according to Ref.~\cite{gengvec} it develops a resonance just above the nominal  ${K^{*}}{\bar K^{*}}$ threshold when the mass distributions of the $K^*$, $\bar{K}^*$ are taken into account to deal with their widths. A manifestation of this resonance was seen in the $J/\psi \rightarrow \eta {K^{*}}^0\bar{K}^{*0}$ decay mode through a peak in the ${K^{*}}{\bar K^{*}}$ mass distribution, which was analyzed in Ref.~\cite{albaxie} and presented as an evidence for this theoretical prediction, although the experiment served to make a more precise determination of the resonance mass. With the aim of suggesting alternative reactions that give extra evidence of this state and serve to put it on firmer grounds, we have investigated the $\eta_c\rightarrow \phi {K^{*}}{\bar K^{*}}$ reaction and have seen that it also provides a shape for the invariant mass distribution of the ${K^{*}}{\bar K^{*}}$ that differs substantially from the one of pure phase space, showing a clear peak around $1830$ MeV, the mass of the  $h_1$ state, while the phase space distribution concentrates its strength at the end of the invariant mass distribution.

We have also evaluated the shape of the same mass distributions starting from an excited state of the $\eta_c$, the $\eta_c (2S)$.  The two distributions are also very different here, and while the one corresponding to the resonance formation peaks again at $1830$ MeV, the one of the phase space keeps growing from threshold up to $2100$ MeV, so the shapes are very distinct. The calculations rely on final state interaction of the $K^* \bar{K}^*$ and one unknown vertex for the bare transition prior to the final state interaction. This does not allow us to determine absolute decay rates, only the shapes of the invariant mass distribution. Yet, one might have a rough estimate of the rates if one sees that the branching ratio for the
$\eta_c\rightarrow \phi {K} \bar{K}$ is $2.9 \times 10^{-3}$~\cite{pdg}. We expect the new rate to be of the same order of magnitude and well in the range of observability. We stressed that the resonance comes from the interaction of ${K^{*}}\bar{K^{*}}$ in single channel which could justify why it has resisted observation for so long, but it would provide a very good example of a clear molecular state coming from the interaction of two vector mesons. At a time when an ongoing discussion on the nature of the different hadronic states is taking place, the establishment of this new resonance would bring important light into this healthy discussion.

\begin{acknowledgments}
X.-L.R acknowledges support from the Innovation Foundation of Beihang University for Ph.D. Graduates. E. Oset acknowledges the hospitality of the School of Physics and Nuclear Energy Engineering of Beihang University.
This work was partly supported by the National
Natural Science Foundation of China under Grants No. 11005007, No. 11375024, and No. 11175002, the Spanish Ministerio de Economia y Competitividad and European FEDER funds under the contract number FIS2011-28853-C02-01, the Generalitat Valenciana in the program Prometeo, 2009/090, and the New Century Excellent Talents in University  Program of Ministry of Education of China under Grant No. NCET-10-0029, the Research Fund  for the Doctoral Program of Higher Education under Grant No. 20110001110087. We acknowledge the support of the European Community-Research Infrastructure Integrating Activity Study of Strongly Interacting Matter (acronym HadronPhysics3, Grant Agreement n. 283286) under the Seventh Framework Program of EU.

\end{acknowledgments}


\end{document}